# DIFFERENTIAL POSTURAL EFFECTS OF PLANTAR-FLEXOR MUSCLES FATIGUE

## UNDER NORMAL, ALTERED AND IMPROVED

## VESTIBULAR AND NECK SOMATOSENSORY CONDITIONS

Nicolas PINSAULT[1] and Nicolas VUILLERME[1*]

[1] Laboratoire TIMC-IMAG, UMR UJF CNRS 5525, La Tronche, France

* Address for correspondence:

Nicolas VUILLERME

Laboratoire TIMC-IMAG, UMR UJF CNRS 5525

Faculté de Médecine

38706 La Tronche cédex

France.

Tel: (33) (0) 4 76 63 74 86

Fax: (33) (0) 4 76 51 86 67

Email: nicolas.vuillerme@imag.fr

**Abstract**

The aim of the present study was to assess the effects of plantar-flexor muscles fatigue on postural control during quiet standing under normal, altered and improved vestibular and neck somatosensory conditions. To address this objective, young male university students were asked to stand upright as still as possible with their eyes closed in two conditions of No Fatigue and Fatigue of the plantar-flexor muscles.

In Experiment 1 (n=15), the postural task was executed in two Neutral head and Head tilted backward postures, recognized to degrade vestibular and neck somatosensory information. In Experiment 2 (n=15), the postural task was executed in two conditions of No tactile and Tactile stimulation of the neck provided by the application of strips of adhesive bandage to the skin over and around the neck.

Centre of foot pressure displacements were recorded using a force platform.

Results showed that (1) the Fatigue condition yielded increased CoP displacements relative to the No Fatigue condition (Experiment 1 and Experiment 2), (2) this destabilizing effect was more accentuated in the Head tilted backward posture than Neutral head posture (Experiment 1) and (3) this destabilizing effect was less accentuated in the condition of Tactile stimulation than that of No tactile stimulation of the neck (Experiment 2).

In the context of the multisensory control of balance, these results suggest an increased reliance on vestibular and neck somatosensory information for controlling posture during quiet standing in condition of altered ankle neuromuscular function.

**Introduction**

In recent years, a growing number of investigations have focused on the effects of plantar-flexor muscles fatigue on postural control during quiet standing. A deteriorated postural control was generally observed following plantar-flexor muscles fatigue, hence stressing the importance of intact ankle neuromuscular function on upright balance control (Ledin et al., 2004; Vuillerme et al., 2002a,b, 2006). Interestingly, this destabilizing effect, reported in the absence of vision, further has been shown to be at least partly alleviated by providing additional visual information (Ledin et al., 2004; Vuillerme et al., 2006), somatosensory cues from the ankle (Vuillerme and Demetz, 2007) and haptic cues from the finger (Vuillerme and Nougier, 2003). These results could reflect a re-weighting of sensory cues in balance control (e.g., Peterka, 2002; Peterka and Loughlin, 2004; Oie et al., 2002; Vuillerme et al., 2001; Vuillerme and Pinsault, 2007), according to which the human postural control system could modulate its dependence upon a given sensory input adaptively to preserve adequate postural control dependent upon current environmental conditions. In conditions of altered ankle neuromuscular function induced by plantar-flexor muscles fatigue, it is thus possible that the central nervous system increases the reliance on visual information (Ledin et al., 2004; Vuillerme et al., 2006), cutaneous inputs from the foot and shank (Vuillerme and Demetz, 2007) and haptic cues from the finger (Vuillerme and Nougier, 2003) to maintain adequate upright stance. Among the sensory inputs relevant to postural control, vestibular and neck somatosensory cues also have been demonstrated to play an important role (e.g. Kavounoudias et al., 1999; Lacour et al., 1997; Peterka and Benolken, 1995; Vuillerme et al., 2005). However, whether the central nervous system increases the reliance on these sensory inputs following plantar-flexor muscles fatigue has not been established yet.

The present study was thus designed to specifically address this issue, by comparing the effects of plantar-flexor muscles fatigue on postural control during quiet standing under normal, altered and improved vestibular and neck somatosensory conditions. We hypothesized that (1) plantar-flexor muscles fatigue would decrease postural control during quiet standing under normal vestibular and neck somatosensory conditions (*hypothesis 1*), (2) this destabilizing effect would be exacerbated under altered vestibular and neck somatosensory conditions (*hypothesis 2*) and (3) this destabilizing effect would be mitigated under improved vestibular and neck somatosensory conditions *(hypothesis 3)*.

**Methods**

Fifteen young male university students (age = 23.3 ± 2.3 years; body weight = 71.0 ± 5.9 kg; height = 179.7 ± 6.1 cm; mean ± S.D.) voluntarily participated in Experiment 1. Fifteen other young male university students (age = 22.7 ± 3.4 years; body weight = 73.3 ± 9.2 kg; height = 177.8 ± 5.7 cm; mean ± S.D.) voluntarily participated Experiment 2. Subjects had to be healthy without any history of injury or pathology to either lower extremity, neck pain, neurological or vestibular impairment, injury or operation in the cervical spine. They gave their informed consent to the experimental procedure as required by the Helsinki declaration (1964) and the local Ethics Committee.

Eyes closed, subjects stood barefoot on the force platform in a natural position (feet abducted at 30°, heels separated by 3 cm), their arms hanging loosely by their sides and were asked to stand as still as possible (Zok et al., 2008). Note that this postural task was executed in the absence of visual information insofar as vision previously was shown to substantially reduce the postural effect caused by plantar-flexor muscles (Ledin et al., 2004; Vuillerme et al., 2006). The eyes closed condition thus avoided visual information interfering with the induced postural behaviours and allowed to isolate/evaluate the specific effect of vestibular and neck somatosensory conditions following localised fatigue of the plantar-flexor muscles on postural control during quiet standing.

------------------------------------

Please insert Figure 1 about here

------------------------------------

In Experiment 1, the postural task was executed in two Neutral head and Head tilted backward postures (Figure 1, upper panel). The Head tilted backward posture is recognised to induce (1) a modification in the orientation of the vestibular organs that may place the utricular otoliths well beyond their working range and render balance related vestibular information unreliable to the central nervous system (e.g., Brandt et al., 1981, 1986; Jackson and Epstein, 1991; Straube et al., 1992) and (2) abnormal sensory inputs arising from neck proprioceptors (e.g. Jackson and Epstein, 1991; Karlberg, 1995; Ryan and Cope, 1955), that represent a challenge for the postural control system. In the Neutral head posture, subjects were asked to keep their head in a straight-ahead direction. In the Head tilted backward posture, they were asked to tilt their head backward for at least 45° in the sagittal plane (e.g., Anand et al., 2002, 2003; Brandt et al., 1981; Buckley et al., 2005; Jackson and Epstein, 1991; Simoneau et al., 1992; Vuillerme and Rougier, 2005). The experimenter always stood by the subjects to monitor their posture and their head position throughout the trial. Subjects were asked to adopt the required posture and to stabilise their body sway. Ten seconds later, the sampling was initiated.

In Experiment 2, the postural task was performed in two conditions of No tactile stimulation and Tactile stimulation of the neck with the head in a neutral position. The No tactile stimulation condition served as a control condition. In the Tactile stimulation condition, four pieces of 3-cm strips of adhesive bandage (Elastoplast, BSN Medical) were applied directly to the skin over and around the neck. The two first strips, starting from external occipital protuberances and ending to the medial angle of scapulas, were positioned directly on the skin over the upper portion of each trapezius muscles. The two last strips, starting from mastoid process of temporal bones and ending on the sternocostoclavicular joint, were applied directly on the skin over each sternocleidomastoid muscles (Figure 1, lower panel). These strips of tape were used to selectively provide cutaneous sensory feedback over and around the neck without adding any mechanical constriction and mechanical pressure on subcutaneous structures (e.g., Simoneau et al., 1997; Vuillerme and Pinsault, 2007). Results of unpublished work performed in our laboratory have indicated that a similar tactile stimulation of the neck to that used in the present experiment has yielded to improved cervical joint position sense in young healthy adults.

For both Experiment 1 and Experiment 2, this experimental procedure was executed the same day before (No Fatigue condition) and after a fatiguing procedure whose aim was to induce a muscular fatigue at the ankle plantar-flexor of both legs until maximal exhaustion (Fatigue condition). As previously done in other studies (Ledin et al., 2004 ; Vuillerme et al., 2006, 2007), standing subjects were asked to perform toe-lifts as many times as possible following the beat of a metronome (40 beats/min). The examiner gave verbal encouragement and checked the exercise performance to ensure that the subjects worked maximally. The fatigue level was reached when subjects were no more able to complete the exercise. Immediately on the cessation of exercise, the subjective exertion level was assessed through the Borg CR-10 scale (Borg, 1990). Subjects rated their perceived fatigue at the plantar-flexors muscles as almost "extremely strong" (mean Borg ratings of 8.3 and 8.5, for Experiment 1 and 2, respectively). The recovery process after fatigue procedures is often considered as a limitation for all fatigue experiments. In the present experiments, to ensure that balance measurement in the Fatigue condition was obtained in a genuine fatigued state, various rules were respected. (1) The fatiguing exercise took place beside the force platform to minimise the time between the exercise-induced fatiguing exercise and the balance measurements (2) the fatiguing exercise was repeated prior to each trial, and (3) the examiner checked that the subjects showed visible signs of fatigue and failed to perform the exercises due to fatigue. In doubtful cases, the examiner encouraged the subject to continue with a few more exercises.

The order of presentation of the two head postures (Experiment 1) and the two conditions of tactile stimulation of the neck (Experiment 2) was randomised over subjects. In Experiment 1, for each condition of No Fatigue and Fatigue of the plantar flexor muscles and each head posture (Neutral and Tilted backward), subjects performed three 32-s trials, for a total of 12 trials. In Experiment 2, for each condition of No Fatigue and Fatigue of the plantar flexor muscles and each condition of tactile stimulation of the neck (No tactile *vs.* Tactile), subjects also performed three 32-s trials, for a total of 12 trials. The sketch of the entire experimental procedure is illustrated in Figure 2 (upper and lower panels for Experiments 1 and 2, respectively).

------------------------------------

Please insert Figure 2 about here

------------------------------------

A force platform (Equi+, model PF01), constituted of an aluminium plate (80 cm each side) laying on three uniaxial load cells, was used to measure the displacements of the centre of foot pressure (CoP) (64 Hz sampling frequency). CoP displacements were processed through a space-time domain analysis including the calculation of the surface area covered by the trajectory of the CoP with a 90% confidence interval (Tagaki et al., 1985) and the variances of positions of the CoP along the medio-lateral (ML) and antero-posterior (AP) axes.

The means of the three trials performed in the each experimental condition were used for statistical analyses. A Kolmogorov-Smirnov test of equality of variances first showed that the distributions used for the analysis did not depart from normality ($Ps > 0.05$).

For Experiment 1, data obtained for the surface area covered by the trajectory of the CoP were then submitted to a 2 Fatigues (No Fatigue *vs.* Fatigue) × 2 Head postures (Neutral *vs.* Tilted backward) analysis of variance (ANOVA) with repeated measures on both factors. To further investigate whether the effects of Fatigue and Head posture were similar according to the ML or AP axes, a 2 Fatigues (No Fatigue *vs.* Fatigue) × 2 Head postures (Neutral *vs.* Tilted backward) × 2 Axes (Medio-lateral *vs.* Antero-posterior) ANOVA with repeated measures on all factors was applied to the variance of the CoP displacements.

For Experiment 2, data obtained for the surface area covered by the trajectory of the CoP were then submitted to a 2 Fatigues (No Fatigue *vs.* Fatigue) × 2 Tactile stimulations of the neck (No tactile *vs.* Tactile) ANOVA with repeated measures on both factors. To further investigate whether the effects of Fatigue and Head posture were similar according to the ML or AP axes, a 2 Fatigues (No Fatigue *vs.* Fatigue) × 2 Tactile stimulations of the neck (No tactile *vs.* Tactile) × 2 Axes (Medio-lateral *vs.* Antero-posterior) ANOVA with repeated measures on all factors was applied to the variance of the CoP displacements.

Post-hoc analyses (*Newman-Keuls*) were performed whenever necessary. Level of significance was set at 0.05.

**Results**

**Experiment 1**

------------------------------------

Please insert Figure 3 about here

------------------------------------

Surface area covered by the trajectory of the CoP

Analysis of the surface area covered by the trajectory of the CoP showed main effects of Fatigue ($F$ (1,14) = 23.92, $P < 0.001$) and Head posture ($F$ (1,14) = 17.54, $P < 0.01$), and a significant two way-interaction of Fatigue × Head posture ($F$ (1,14) = 5.72, $P < 0.05$) (Figure **3**).

The decomposition of this interaction into its simple main effects indicated that (1) the Fatigue condition yielded increased CoP surface area relative to the No Fatigue condition in the Neutral head posture ($P < 0.05$), and (2) this effect was more accentuated in the Head tilted backward posture ($P < 0.001$).

------------------------------------

Please insert Figure 4 about here

------------------------------------

Variance of the CoP displacements along the ML and AP axes

Analysis of the variance of the CoP displacements showed main effects of Fatigue ($F$ (1,14) = 20.14, $P < 0.001$), Head posture ($F$ (1,14) = 16.50, $P < 0.01$) and Axis ($F$ (1,14) = 34.98, $P < 0.001$), three significant two-way interactions of Fatigue × Head posture ($F$ (1,14) = 6.17, $P < 0.05$), Fatigue × Axis ($F$(1,14) = 16.18, $P<0.01$) and Head posture × Axis ($F$ (1,14) = 7.59, $P < 0.05$), and a significant three-way interaction of Fatigue × Head posture × Axis ($F$ (1,14) = 8.72, $P < 0.05$) (Figure 4).

The decomposition of the two-way interactions of Fatigue × Head posture, Fatigue × Axis and Head posture × Axis indicated that (1) the Fatigue condition yielded larger increased variance of the CoP displacements relative to the No Fatigue condition in the Head tilted backward ($P < 0.001$) than Neutral head posture ($P < 0.05$), (2) the Fatigue condition yielded larger increased variance of the CoP displacements relative to the No Fatigue condition along the AP ($P < 0.001$) than ML axis ($P < 0.05$), and (3) the Head tilted backward posture yielded larger increased variance of the CoP displacements relative to the Neutral head posture along the AP ($P < 0.001$) than ML axis ($P < 0.05$), respectively.

The decomposition of the three-way interaction of Fatigue × Head posture × Axis further indicated that the destabilizing effect of Fatigue observed in the Head tilted backward posture was larger along the AP ($P < 0.001$) than ML axis ($P < 0.01$).

## Experiment 2

-------------------------------------

Please insert Figure 5 about here

-------------------------------------

### Surface area covered by the trajectory of the CoP

Analysis of the surface area covered by the trajectory of the CoP showed main effects of Fatigue ($F$ (1,14) = 9.69, $P < 0.01$) and Tactile stimulation ($F$ (1,14) = 7.19, $P < 0.05$), and a significant two way-interaction of Fatigue × Tactile stimulation ($F$ (1,14) = 5.58, $P < 0.05$) (Figure 5).

The decomposition of this interaction into its simple main effects indicated that (1) the Fatigue condition yielded increased CoP surface area relative to the No Fatigue condition in the No tactile stimulation condition ($P < 0.001$), and (2) this effect was less accentuated in the Tactile stimulation condition ($P < 0.01$).

-------------------------------------

Please insert Figure 6 about here

-------------------------------------

### Variance of the CoP displacements along the ML and AP axes

Analysis of the variance of the CoP displacements showed main effects of Fatigue ($F$ (1,14) = 10.88, $P < 0.01$), Tactile stimulation ($F$ (1,14) = 4.61, $P < 0.05$) and Axis ($F$ (1,14) =

10.04, $P < 0.01$), two significant two-way interactions of Fatigue × Tactile stimulation ($F$ (1,14) = 6.02, $P < 0.05$) and Fatigue × Axis ($F$ (1,14) = 8.75, $P<0.05$), and a significant three-way interaction of Fatigue × Tactile stimulation × Axis ($F$ (1,14) = 4.83, $P < 0.05$) (Figure 6).

The decomposition of the two-way interactions of Fatigue × Tactile stimulation and Fatigue × Axis indicated that (1) the Fatigue condition yielded smaller increased variance of the CoP displacements relative to the No Fatigue condition in the Tactile stimulation ($P < 0.01$) than No tactile stimulation condition ($P < 0.001$), and (2) the Fatigue condition yielded increased variance of the CoP displacements relative to the No Fatigue condition along the AP ($P < 0.001$), whereas no significant difference was observed along than ML axis ($P > 0.05$), respectively.

The decomposition of the three-way interaction of Fatigue × Tactile stimulation × Axis further indicated that the destabilizing effect of Fatigue observed along the AP axis was smaller in the Tactile stimulation ($P < 0.001$) than No tactile stimulation condition ($P < 0.01$).

**Discussion**

The aim of the present study was to assess the effects of plantar-flexor muscles fatigue on postural control during quiet standing under normal, altered and improved vestibular and neck somatosensory conditions. To address this objective, young male university students were asked to stand upright as still as possible with their eyes closed in two conditions of No Fatigue and Fatigue of the plantar-flexor muscles.

In Experiment 1 (n=15), the postural task was executed in two Neutral head and Head tilted backward postures, recognized to degrade vestibular and neck somatosensory information. In Experiment 2 (n=15), the postural task was executed in two conditions of No tactile and Tactile stimulation of the neck, provided by the application of strips of adhesive bandage to the skin over and around the neck.

Centre of foot pressure displacements were recorded using a force platform.

Under normal vestibular and neck somatosensory condition, plantar-flexor muscles fatigue worsened postural control during quiet standing, as indicated by the increased surface area covered by the trajectory of the CoP observed in the Fatigue relative to the No Fatigue condition (Neutral head posture condition in Experiment 1, Figure 3, left part and No tactile stimulation of the neck condition in Experiment 2, Figure 5, left part). Analysis of the variance of the CoP displacements further indicated that the destabilizing effect of Fatigue was more accentuated along the AP than ML axis (Figures 4 and 6). These results, in accordance with previous reports (Ledin et al., 2004; Vuillerme et al., 2006), support our hypothesis 1. If one considers postural control mainly as a perceptual-motor process (Schmidt, 1975), these results could stem from an alteration of both the sensory (e.g., Vuillerme et al., 2007; Vuillerme and Boisgontier, 2008) and the motor sides (e.g., Trappe et al., 2001) of this process, caused by the fatigue of the plantar-flexor muscles. On the one hand, degraded senses of position (Vuillerme et al., 2007) and force (Vuillerme and Boisgontier, 2008) at the ankle joint have recently been reported following plantar-flexor muscles fatigue. On the other hand, the observations of impaired postural control only when individuals could not used visual sensory cues (Ledin et al., 2004; Vuillerme et al., 2006), somatosensory cues from the ankle (Vuillerme and Demetz, 2007) and haptic cues from the finger (Vuillerme and Nougier, 2003), could suggest that fatigue-induced plantar-flexor muscles weakness (Trappe et al., 2001), *per se*, could not have caused the increased CoP displacements observed in the Fatigue condition. Although a similar thought process/interpretation recently has been given by Bulter et al. (2008), further investigations are nevertheless necessary to address this issue. Although these two non exclusive propositions are supported by larger destabilising effects observed along the AP than ML axis, when considering what the fatiguing exercise involved in terms of joints and tendons receptors stimulation and muscles recruitment (i.e. ankle plantar-flexors muscles), another explanation for subjects' increased postural sway in the fatigue condition could be the increased ventilation rate induced by muscular exercise. At this point, we would like to mention that respiration rate was not measured quantitatively. To address this issue, as previously done by Davidson et al. (2004), a small control experiment was performed on three subjects to (1) quantify the increase in respiration rate due to the fatiguing exercise, and (2) determine if an identical increase in respiration rate in the absence of plantar-flexors muscles increases CoP displacements. For these 3 subjects, respiration rate was measured before and following the fatiguing exercise. An increased respiration rate (+ 3 breaths/min) was observed following the fatiguing exercise. These three subjects were asked to stand as still as possible while breathing at their respiration rate measured following the fatiguing exercise. Results showed that the increased in respiration rate in absence of plantar-flexor muscles fatigue induced weak but non significant effects on CoP displacements ($P>0.05$). This suggests that the increase in respiration rate in our study resulting from the fatiguing exercise did not contribute significantly to the large increase in CoP displacements reported in the Fatigue condition. Note that similar conclusions were made by Nardone et al. (1997) and Davidson et al. (2004).

When vestibular and somatosensory information from the neck was altered by asking subject to adopt a head tilted backward posture (e.g., Brandt et al., 1981, 1986; Jackson and Epstein, 1991; Karlberg, 1995; Ryan and Cope, 1955; Straube et al., 1992) (Experiment 1), the destabilizing effect of plantar-flexor muscles fatigue was more accentuated, as indicated by the significant two-way interaction of Fatigue × Head posture observed for the surface area covered by the trajectory of the CoP (Figure 3). This result supports our hypothesis 2. Furthermore, the observation of a significant three-way interaction of Fatigue × Head posture × Axis for the variance of the CoP displacements indicated that the destabilizing effect of Fatigue observed in the Head tilted backward head posture was more accentuated along the AP than ML axis (Figure 4). This result can be explained by the fact that the fatiguing exercise applied to the ankle plantar-flexor muscles and the head extended posture have been shown to separately induce larger destabilising effects along the AP than ML axis (Ledin et al., 2004; Vuillerme et al., 2006 ; Vuillerme and Rougier, 2005).

When somatosensory information from the neck was improved by increasing cutaneous feedback through the application of strips of adhesive bandage to the skin over and around the neck (e.g., Simoneau et al., 1997; Vuillerme and Pinsault, 2007) (Experiment 2), the destabilizing effect of plantar-flexor muscles fatigue was mitigated, as indicated by the significant two-way interaction of Fatigue × Tactile stimulation observed for the surface area covered by the trajectory of the CoP (Figure 5). This result supports our hypothesis 3. Furthermore, the observation of a significant three-way interaction of Fatigue × Tactile stimulation × Axis for the variance of the CoP displacements indicated that the stabilizing effect of Tactile stimulation observed in the Fatigue condition was more accentuated along the AP than ML axis (Figure 6). Considering (1) results of unpublished work performed in our laboratory reporting the ability of young healthy adults to take advantage of increased neck cutaneous information provided by the by strips of adhesive bandage applied to the neck to improve their cervical joint position sense, and (2) the relationship existing between cervical joint position sense and postural control during quiet standing (e.g., Treleaven et al., 2006; 2008), it is possible that the stabilizing effect of tactile stimulation under fatigue condition (Experiment 2, Figures 5 and 6), stem from an increased cervical proprioceptive acuity induced by the tactile stimulation of the neck. Note that similar postural control improvements have previously been reported when tactile stimulation was provided at other joints, such as hip (McNair and Heine, 1999), knee (Birmingham et al., 2001; Hassan et al., 2001; Kaminski and Perrin, 1996), or ankle (Vuillerme and Pinsault, 2007; Vaillant et al., 2008).

On the whole, results of both experiments 1 and 2 suggest an increased reliance on vestibular and neck somatosensory information for controlling posture during quiet standing in condition of altered ankle neuromuscular function induced by plantar-flexor muscle fatigue. These results could lend support to the hypothesis of the sensory re-weighting hypothesis of human postural control (e.g. Peterka, 2002; Peterka and Loughlin, 2004; Oie et al., 2002; Vuillerme et al., 2001, 2005; Vuillerme and Pinsault, 2007). Sensory reweighting mechanism, first suggested by Nashner and colleagues more than 30 years ago (e.g., see Nashner 1976; Nashner et al. 1982), is recognised as a crucial component of postural control. It supposes that he central nervous system is able to dynamically and selectively adjust the relative contributions of sensory inputs (i.e., the sensory weights) to control balance depending on the sensory contexts and the neuromuscular constraints acting on the subject. For instance, previous studies have reported an increased reliance on vestibular and/or neck somatosensory cues when somatosensory information from the support surface was altered by standing either on a compliant (e.g., Buckley et al., 2005; Horak and Hlavacka, 2001; Vuillerme et al., 2005, 2008), sway-referenced (e.g., Cenciarini and Peterka, 2006; MacDougall et al., 2006), unstable (e.g. Fitzpatrick et al., 1994; Ivanenko et al., 1999) or moving support surface (e.g. Hlavacka and Horak, 2006; Inglis et al., 1995), and when proprioceptive information from the ankles was degraded by vibratory stimulation (e.g., Dzurková and Hlavacka, 2007; Fransson et al., 2007; Hlavacka et al., 1995; Kavounoudias et al., 1999; Vuillerme and Cuisinier, 2008). Our results extend this observation to a more normal upright posture and to situations everyone could encountered in which afferent signals and efferent outputs in lower-limb muscles are impaired by a fatiguing exercise, that is in conditions everyone could encountered in sportive, occupational or daily activities. As above-mentioned, such a sensory re-weighting in postural control following plantar-flexor muscles fatigue is not limited to vestibular and neck somatosensory inputs. Indeed, previous studies have evidenced the ability of the central nervous system to up-weight visual information (Ledin et al. 2004; Vuillerme et al., 2006), cutaneous inputs from the foot and shank (Vuillerme and Demetz, 2007) and haptic cues from the finger (Vuillerme and Nougier, 2003), providing more accurate and reliable information for ensuring adequate postural control in conditions of plantar-flexor muscles fatigue.

Finally, we believe that the present findings could have implications in clinical and rehabilitative areas. Indeed, while results of Experiment 2 evidenced that the effects of plantar-flexor muscles fatigue on postural control were mitigated in conditions of improved neck somatosensory information provided by the application of strips of adhesive bandage to the skin over and around the neck, whether individuals could benefit from an artificial head position-based biofeedback to preserve (Vuillerme and Cuisinier, 2008; Vuillerme et al., 2008) postural control in conditions of plantar-flexor muscles fatigue is currently being investigated.

**Acknowledgements**

This work was partially supported by MENRT-France. The authors would like to thank subject volunteers and the anonymous reviewers for helpful comments and suggestions. Special thanks also are extended to R. Murphy, M. Gillet and R. Bertrand for their help in data collection and M. Harcelle for various contributions.

**Figure captions**

**Figure 1.** Schematic representation of (1) the head postures adopted in two Neutral head and Head tilted backward postures in Experiment 1 (upper panel) and (2) the application of adhesive bandage strips over and around the neck proposed in the two conditions of No tactile stimulation and Tactile stimulation of the neck in Experiment 2 (lower panel).

**Figure 2.** Schematic representation of the experimental procedure used on the present study. The order of presentation of two Neutral head and Head tilted backward postures (Experiment 1, upper panel) and the two conditions of No tactile stimulation and Tactile stimulation of the neck (Experiment 2, lower panel) was randomised over subjects. Each experimental condition consisted in $3 \times 32$ seconds trials with 1 minute rest between trials. The Fatigue condition was executed 5 minutes after the No Fatigue condition.

**Figure 3.** Mean and standard error of mean of the surface area covered by the trajectory of the CoP obtained in the two Neutral head and Head tilted backward postures and the two conditions of No Fatigue and Fatigue of the plantar-flexor muscles. The two conditions of No Fatigue and Fatigue are presented with different symbols: No Fatigue (*white bars*) and Fatigue (*black bars*).

**Figure 4.** Mean and standard error of mean of the variance of the CoP displacements along the medio-lateral (ML) and antero-posterior (AP) axes obtained in the two Neutral and Head tilted backward postures and the two conditions of No Fatigue and Fatigue of the plantar-flexor muscles. The two conditions of No Fatigue and Fatigue are presented with different symbols: No Fatigue (*white bars*) and Fatigue (*black bars*).

**Figure 5.** Mean and standard error of mean of the surface area covered by the trajectory of the CoP obtained in the two conditions of No tactile stimulation and Tactile stimulation of the neck and the two conditions of No Fatigue and Fatigue of the plantar-flexor muscles. The two conditions of No Fatigue and Fatigue are presented with different symbols: No Fatigue (*white bars*) and Fatigue (*black bars*).

**Figure 6.** Mean and standard error of mean of the variance of the CoP displacements along the medio-lateral (ML) and antero-posterior (AP) axes obtained in the two conditions of No tactile stimulation and Tactile stimulation of the neck and the two conditions of No Fatigue and Fatigue of the plantar-flexor muscles. The two conditions of No Fatigue and Fatigue are presented with different symbols: No Fatigue (*white bars*) and Fatigue (*black bars*).

**Figure 1**

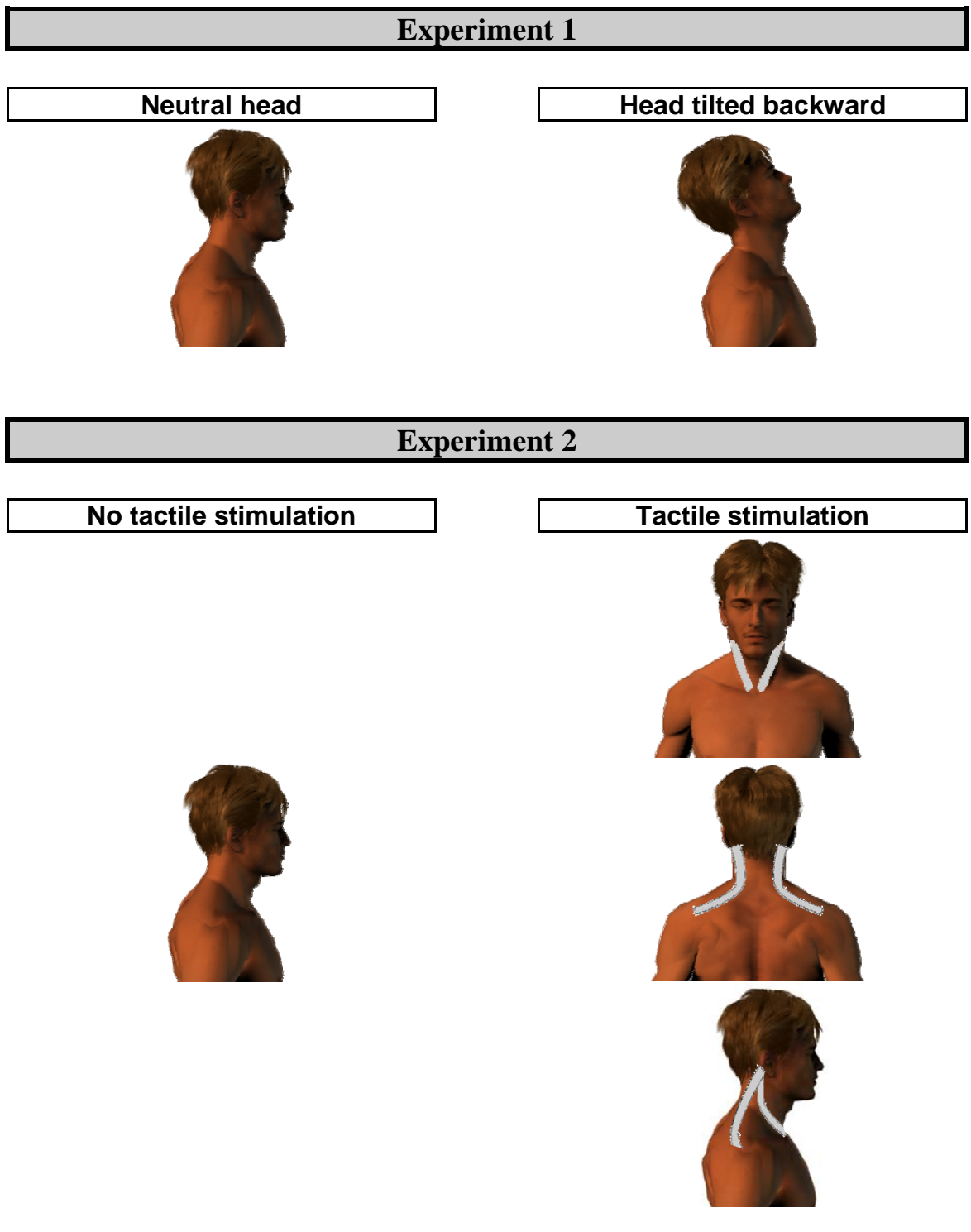

**Figure 2**

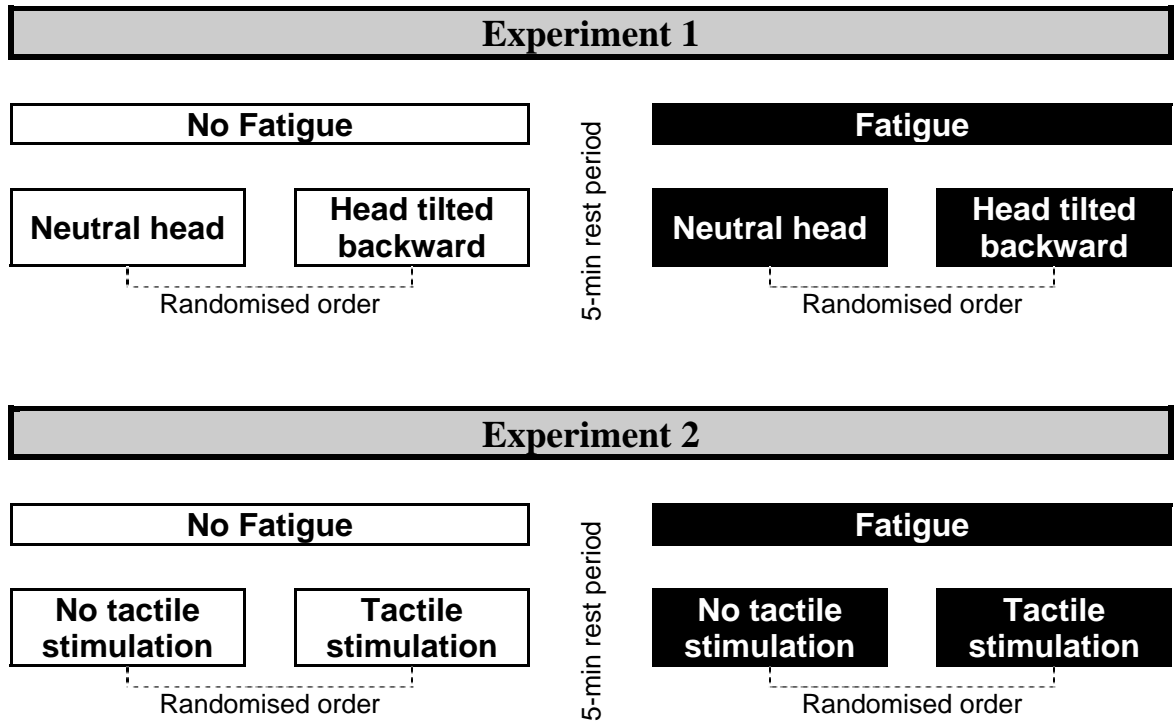

**Figure 3**

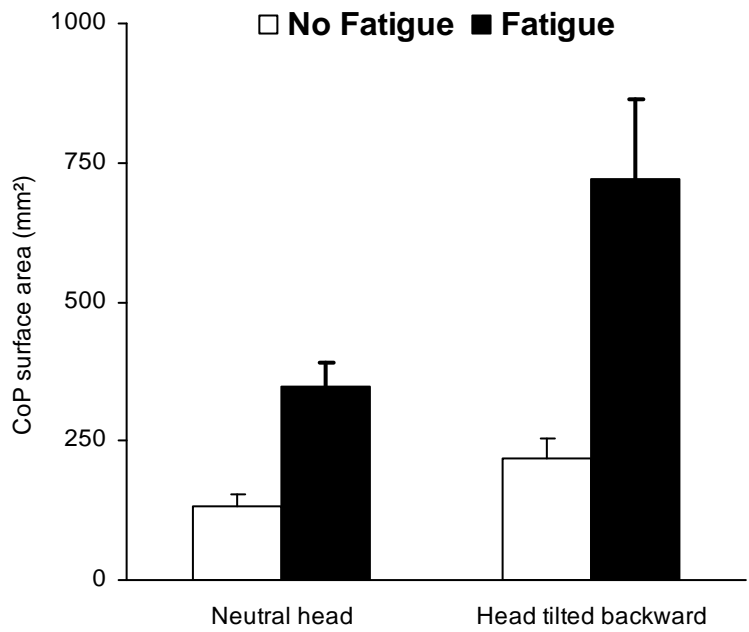

**Figure 4**

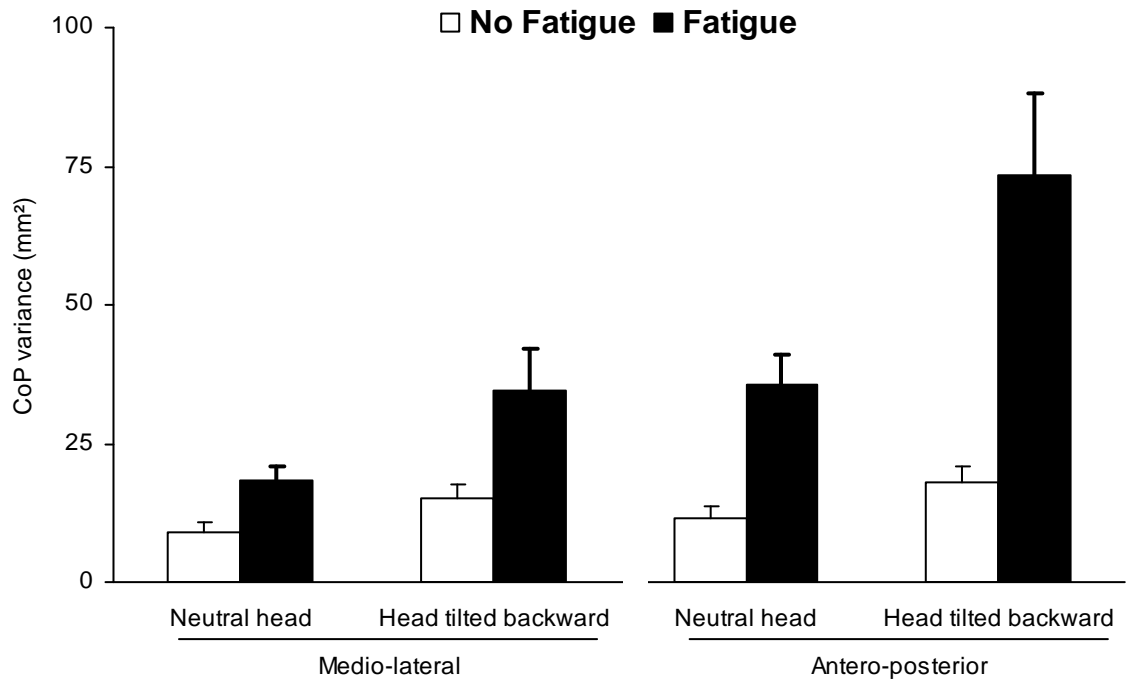

**Figure 5**

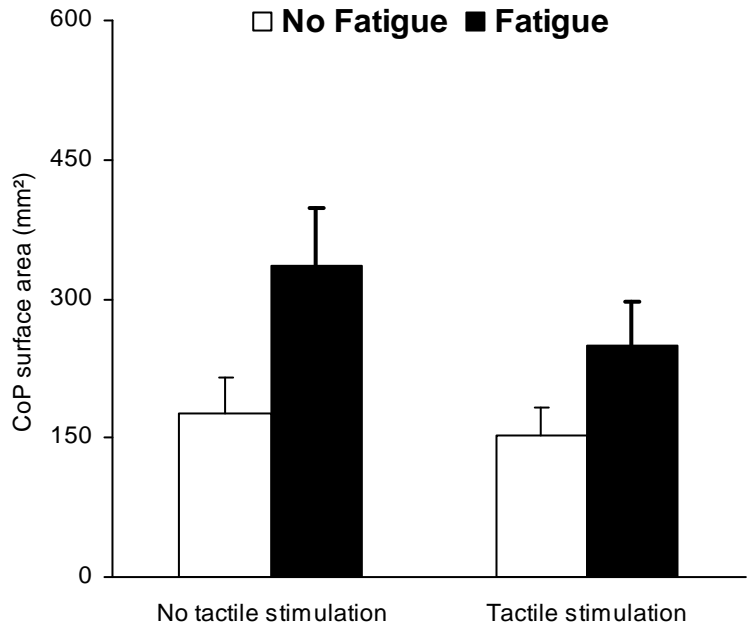

**Figure 6**

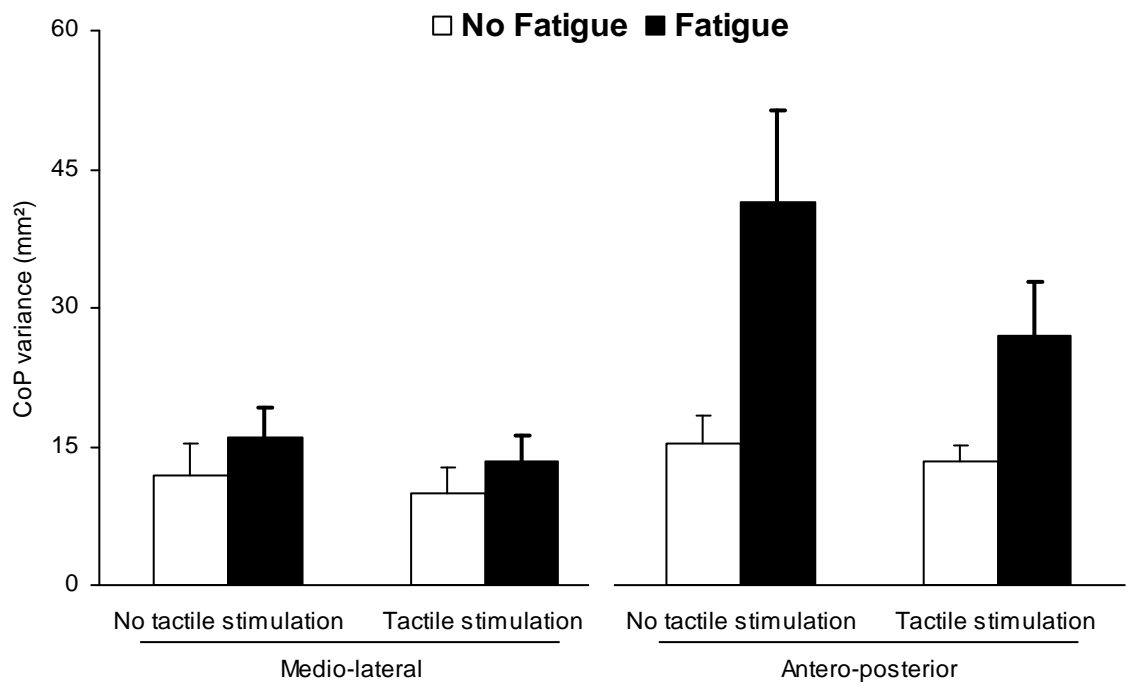